\documentclass[reprint,
nofootinbib,
nobibnotes,
 amsmath,amssymb
aps,
]{revtex4-1}
\setcitestyle{authoryear,round}
\usepackage{graphicx}
\usepackage{dcolumn}
\usepackage{bm}
\usepackage{outlines}
\usepackage{xcolor, soul}

\usepackage{hyperref}
\hypersetup{colorlinks=true,linkcolor=blue,citecolor=blue,filecolor=blue,urlcolor=blue}

\graphicspath{ {./figures/} }

\begin{document}

\preprint{APS/123-QED}

\title{Constraining neutrino mass with tomographic weak lensing peak counts}

\author{Zack Li}
 \email{zequnl@astro.princeton.edu}

\author{Jia Liu}
\affiliation{ Department of Astrophysical Sciences, Princeton University, Princeton, NJ 08544, USA 
}

\author{Jos\'{e} Manuel Zorrilla Matilla}
\affiliation{Department of Astronomy, Columbia University,
New York, NY 10027, USA}

\author{William R. Coulton}
\affiliation{Institute of Astronomy and Kavli Institute for Cosmology Cambridge, Madingley Road, Cambridge, CB3 0HA, UK}
\affiliation{Joseph Henry Laboratories, Princeton University, Princeton, NJ 08544, USA}

\date{\today}

\begin{abstract}
Massive cosmic neutrinos change the structure formation history by suppressing perturbations on small scales. Weak lensing data from galaxy surveys probe the structure evolution and thereby can be used to constrain the total mass of the three active neutrinos. However, much of the information is at small scales where the dynamics are nonlinear. Traditional approaches with second order statistics thus fail to fully extract the information in the lensing field. In this paper, we study constraints on the neutrino mass sum using lensing peak counts, a statistic which captures non-Gaussian information beyond the second order. We use the ray-traced weak lensing mocks from the Cosmological Massive Neutrino Simulations (\texttt{MassiveNuS}), and apply LSST-like noise. We discuss the effects of redshift tomography, multipole cutoff $\ell_{\rm max}$ for the power spectrum, smoothing scale for the peak counts, and constraints from peaks of different heights. We find that combining peak counts with the traditional lensing power spectrum can improve the constraint on neutrino mass sum, $\Omega_m$, and $A_s$ by 39\%, 32\%, and 60\%, respectively, over that from the power spectrum alone.
\end{abstract}

\maketitle

\section{\label{sec:introduction}Introduction}

While ground-based neutrino oscillation experiments established the mass sum of the three active neutrinos to be nonzero~\citep{becker-szendy1992,fukuda1998,Ahmed2004}, the strongest constraint on the upper bound currently comes from cosmology. With nonzero masses, cosmic neutrinos can change the expansion history and suppress the growth of structure.
As such, the combined measurements of the cosmic microwave background~(CMB), CMB lensing, and baryon acoustic oscillation put an upper limit of 0.12~eV on the neutrino mass sum~\citep{planck2018}, under the assumption of a $\Lambda$--Cold Dark Matter~($\Lambda$CDM) cosmology.

A promising way to improve upon the current constraints is the inclusion of weak lensing of galaxies, which measures the distortion of background galaxies by foreground matter. Compared with CMB lensing, weak lensing of galaxies probes the structure growth at lower redshifts. Combined together, the primary CMB, CMB lensing, and weak lensing capture the evolution of structure growth under the influence of massive neutrinos from when they were relativistic ($z\ge$ a few $\times$ 100) to non-relativistic today. Pioneering weak lensing surveys such as the Canada-France-Hawaii Telescope Lensing Survey~(CFHTLenS)~\citep{Heymans2012}, Kilo-Degree Survey~(KiDS)~\citep{Hildebrandt2017}, Dark Energy Survey~(DES)~\citep{2017DES}, and Hyper Suprime-Cam (HSC)~\citep{Mandelbaum2017} have already demonstrated the power of weak lensing. Weak lensing is also a key science component of next generation surveys, including the \texttt{LSST}\footnote{Large Synoptic Survey Telescope: \url{http://www.lsst.org}}, WFIRST\footnote{Wide-Field Infrared Survey Telescope: \url{http://wfirst.gsfc.nasa.gov} }, and Euclid\footnote{Euclid: \url{http://sci.esa.int/euclid} }. 

Massive neutrinos affect the growth of structure most prominently on small scales. However, nonlinear growth dominates these scales at low redshifts. 
The usual techniques to quantify weak lensing observables---the two-point correlation function or its Fourier transformation, the power spectrum---are insufficient in capturing all the nonlinear information.
Additional non-Gaussian information can be extracted by including higher-order statistics, such as the three-point function or the bispectrum~\citep{Takada2003,Takada2004,DZ05,Sefusatti2006,Berge2010,Vafaei2010,Fu2014}. 
However, computing higher-order correlation functions quickly becomes prohibitive as the number of possible shapes increases logarithmically and the signal-to-noise degrades.

Counting peaks in weak lensing convergence~($\kappa$) maps has been proposed to be one simple yet efficient non-Gaussian statistic~\citep{Jain2000b,Marian2009,Maturi2010,Marian+2013,Lin&Kilbinger2015a,Lin&Kilbinger2015b,LiuHaiman2016}.
Previous theoretical works have found up to a factor of two improvement in constraining cosmological parameters such as $\Omega_m$, $\sigma_8$, and $w$, when peak counts and two-point statistics are combined, compared with using the latter alone~\citep{Yang2011}. This prediction is recently supported by measurements using observational data from the CFHTLenS~\citep{Liu2015}, CS82~\citep{Liux2015}, DES~\citep{Kacprzak2016}, and KiDS~\citep{shan2018, martinet2018} surveys. The neutrino mass sum, however, has been typically assumed to be zero in these works. 
\citet{Peel2018} is the first to study peak counts in massive neutrino cosmology. They found that peak counts in aperture mass maps can help break the degeneracy between neutrino mass and modified gravity, and that peaks outperform the third- and fourth-order moments. 

In this paper, we extend the study of peak counts to cosmological constraints including the neutrino mass sum, and forecast for an \texttt{LSST}-like survey. Because the linear theory breaks down at small scales where nonlinear evolution dominates, we model the lensing power spectrum and peak counts using N-body ray-tracing simulations. In particular, we use the Cosmological Massive Neutrino Simulations ~\citep[\texttt{MassiveNuS},][]{liu2018}, which includes a set of 101 cosmological models, all with different neutrino masses, and other two varying parameters: the matter density $\Omega_m$ and the primordial power spectrum amplitude $A_s$. We use an \texttt{LSST}-like noise and redshift distribution for five tomographic redshift bins, 
and study the constraints from the power spectrum and peak counts separately and jointly. Our companion papers explore the constraints on the neutrino mass sum from other non-Gaussian statistics, including the bispectrum~(Coulton et al. in prep), one-point probability distribution function~\citep{LiuMadhavacheril2018}, and Minkowski Functionals~(Marques et al. in prep). In addition, \citet{Kreisch2018} examined the effect of massive neutrinos on cosmic voids.

We begin this paper with theoretical background in Section~\ref{sec:background}. We describe the \texttt{MassiveNuS} simulations and lensing maps used in this work in Section~\ref{sec:simulations}. In Section~\ref{sec:analysis} we describe our analysis process, including the lensing power spectrum, peak counts, and the likelihood function. In Section \ref{sec:results} we  forecast the cosmological constraints given expected \texttt{LSST} galaxy density and shape noise. 
We summarize our findings and discuss implications in Section \ref{sec:conclusion}.

\section{\label{sec:background}Background}

\subsection{Effects of massive cosmic neutrinos}

Big bang cosmology predicts a relic sea of cosmic neutrinos~\citep{lesgourgues2006}, which is coupled to the quark-baryon fluid at early times and decouples at a temperature of $T_{\rm dec} \approx 1$ MeV. Massive neutrinos contribute to the expansion of the universe but not to gravitational clustering below their free-streaming scale, and hence slow down the growth of matter perturbations. During matter domination, the linear theory solution to the growth of matter perturbation---$\delta_{\rm cdm}\propto a$ for zero neutrino masses---is altered by massive neutrinos,
\begin{equation}
\delta_{\rm cdm} \propto a^{1 - \frac{3}{5} f_{\nu}},
\end{equation}
where we assume the neutrino to matter density fraction $f_{\nu} \equiv \Omega_{\nu} / \Omega_m\ll1$. Measurements of the matter perturbation therefore enable us to measure the total mass of neutrinos.

At present, only the differences between the squared masses of the three species are known, $\Delta m_{21}^2 \equiv m_2^2 - m_1^2 = 7.37^{+0.60}_{-0.44} \times 10^{-5}$ eV$^2$ and $|\Delta m^2| \equiv |m_3^2 - (m_1^2 + m_2^2)/2| = 2.5^{+0.13}_{-0.13} \times 10^{-3}$ eV$^2$ from oscillation experiments~\citep{patrignani2016}. As a result, there are two possible ways to arrange the three neutrino masses, the normal hierarchy~($m_3>m_1,m_2$) and the inverted hierarchy~($m_1,m_2>m_3$) with minimum total masses of $\sim 0.06$ eV and $\sim 0.1$ eV, respectively.

\subsection{Weak lensing}
Gravitational lensing describes the phenomenon where photons from distant objects are deflected by matter before reaching the observer, resulting in a distorted image of the object. The image, usually of a galaxy, can change in brightness and shape as the result of lensing, with the degree of modification in proportion to the mass of the structure inducing the deflection. In strong lensing, these image distortions can be very dramatic, though such events are rare. All galaxies experience weak lensing---minute distortions by foreground matter (``lens''). The effect of weak lensing is difficult to measure on individual galaxies, due to the large uncertainties in their intrinsic brightness and shapes. Nevertheless, robust signals can be achieved statistically by averaging over a large sample of galaxies. This method holds promise for its potential to probe small scale structures and sensitivity to all matter.

The lens creates a deflection angle $\vec{\alpha}$ between the source position $\vec{\beta}$ and the observed position $\vec{\theta}$ ~\citep{bartelmann2001},
\begin{align}
\vec{\theta} = \vec{\beta} + \vec{\alpha}(\vec{\theta}).
\end{align}
For a thin lens, the deflection angle is related to the lensing potential~$\Psi$, through
\begin{align}
\vec{\alpha}=&\nabla \Psi\\
\Psi(\vec{\theta}) =& \frac{D_{ds}}{D_d D_s} \frac{2}{c^2} \int_0^{z_s} \Phi( D_d \vec{\theta}, z)\,dz,
\end{align}
where $D_s$, $D_d$, $D_{ds}$ are the angular diameter distances between the source and observer, lens and observer, and source and lens, respectively. $\Phi$ is the three-dimensional (3D) gravitational potential. The weak lensing convergence $\kappa$ is a projected density, and its relation to the lensing potential,
\begin{align}
\kappa=\frac{1}{2}\nabla^2 \Psi,
\end{align}
can be viewed as the 2D analogy of the Poisson equation linking the density field $\rho$ to the 3D potential $\Phi$.

\subsection{Lensing power spectrum}
The power spectrum of the convergence map $C_{\ell}$ is a weighted projection of the 3D matter power spectrum $P(k,z)$. In a flat universe, under the Born and flat sky approximations, 
\begin{align}C_{\ell} &= \int_0^{z_s} \frac{H(z)}{c\chi^2(z)} W^2(z) P\left( k = \frac{l}{\chi(z)},\,z \right)dz\\
W(z) &= \frac{3}{2} \Omega_m H_0^2 \frac{1+z}{H(z)} \frac{\chi(z)}{c} \\ &\times \int_z^{z_s}  \frac{dn(z_s)}{dz_s} \frac{\chi(z_s) - \chi(z)}{\chi(z_s)}dz_s.
\end{align}
where the $\chi(z)$ is the comoving distance, $H$ is the Hubble parameter with a present day value $H_0$, $z_s$ is the source redshift, and $dn/dz_s$ is the source redshift distribution.

\subsection{Lensing peak counts}
\label{sec:background_peaks}
Lensing peaks are the local maxima in a convergence $\kappa$ map.  To remove noises that are dominating the small scales, the maps are first smoothed, typically with a Gaussian filter. The optimal smoothing scale depends on the galaxy number density and the shape noise, and we found that in our case 2 arcmin is optimal. In a pixelized map, peaks are the pixels with values higher than all the neighboring pixels. They are quantified as a function of height, i.e.\ the pixel's $\kappa$ value, or the detection significance to noise (S/N). They are particularly sensitive to nonlinear structures, and hence provide information beyond second order statistics like the power spectrum.

High significance peaks (S/N$>$3) are typically the result of single, massive halos, while medium and low peaks come from aligned smaller halos along the line of sight~\citep{Yang2011,LiuHaiman2016}. To date, (semi-)analytic models are available only for the high peaks~\citep{Fan2010,Lin&Kilbinger2015a}, and medium to low peaks are modeled using N-body simulations. While high peaks have high signal and are commonly used in cluster searches, medium and low peaks have been shown to contain comparable cosmological information~\citep{Yang2011}, a feature that we also explore in this paper. We include peaks with a wide range of heights with S/N=[$-0.6$, 6].

\section{Simulations} \label{sec:simulations}

\begin{figure*}
  \includegraphics[width=1.8\columnwidth]{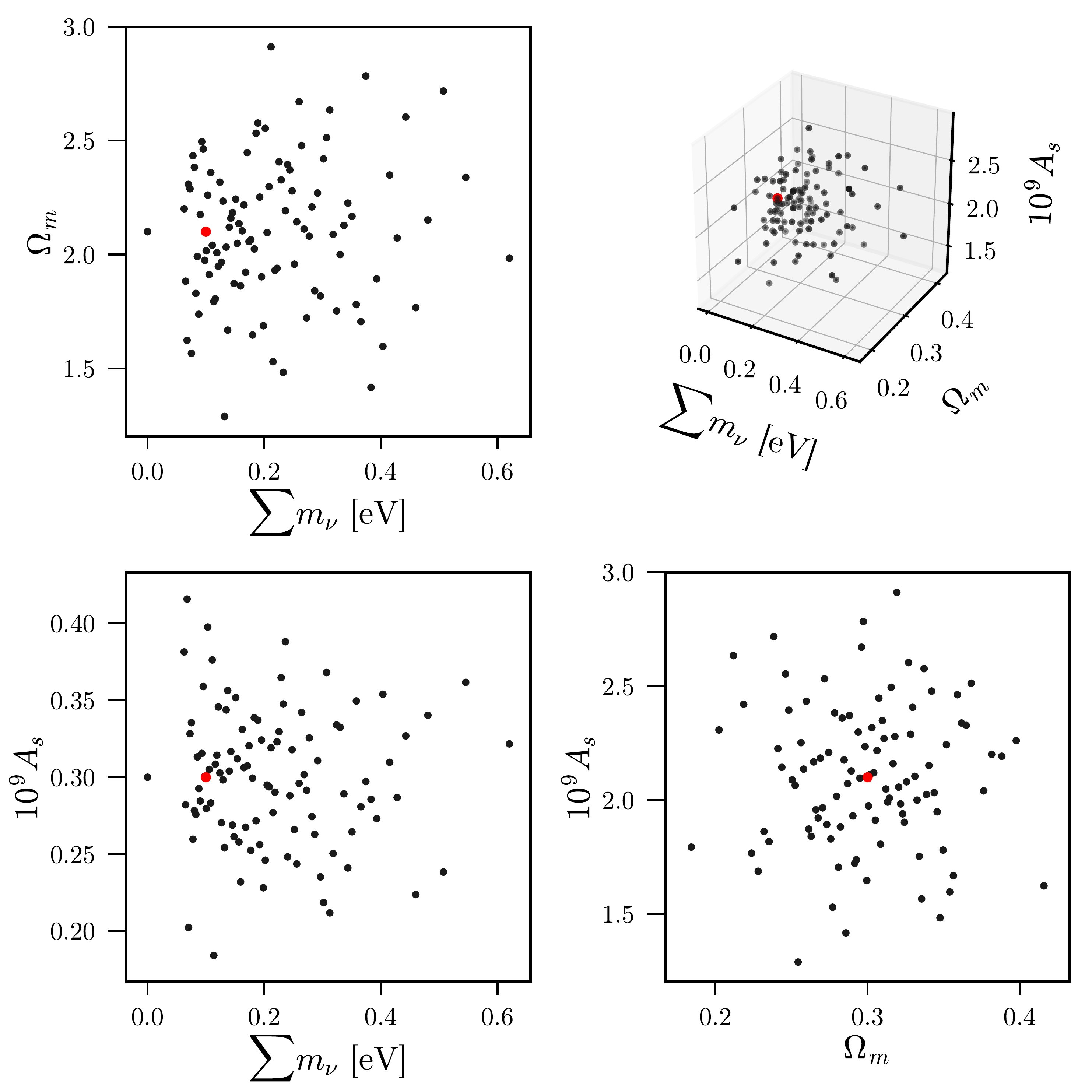}
  \caption{\label{fig:sim_parameter_distribution}The models (black) used in the \texttt{MassiveNuS} simulations, projected onto 2D planes. The fiducial model $\sum m_{\nu} = 0.1$ eV, $\Omega_m = 0.3$, $A_s = 2.1 \times 10^{-9}$ is highlighted in red.}
\end{figure*}

We use the public \texttt{MassiveNuS} simulation suite~\citep{liu2018}, a set of 101 N-body simulations with varying $\sum m_{\nu}$, the total mass of massive neutrinos, $\Omega_m = \Omega_c + \Omega_b + \Omega_{\nu}$, the present matter-component density, and $A_s$, the primordial power spectrum normalization at pivot scale $k_0 = 0.05$ Mpc$^{-1}$. The model sampling of the simulation is shown in Fig.~\ref{fig:sim_parameter_distribution}. The simulations have a 512 Mpc/$h$ box size and $1024^3$ particles. 

The linear power spectrum at $z=99$ was first computed using the Boltzmann code CAMB~\citep{Lewis:1999bs, Howlett:2012mh}. The initial conditions were generated using a modified version of N-GenIC~\citep{springel2005}, which includes the radiation contribution as well as massive neutrinos.  The N-body simulations were run with a modified version of Gadget-2~\citep{springel2005}, which includes a background density of neutrinos using a linear response algorithm~\citep{ali-hamoud2013,bird2018}. 

Lensing convergence maps  are generated using the public ray-tracing code LensTools~\citep{petri2016}. Each snapshot is cut into four equal-sized slices and CDM particles are projected onto a 2D density plane. They are then combined with the neutrino density plane. Light rays are shot from the center of the $z$=0 plane backwards through time, and the deflection angles are computed at each plane. The trajectories of the light rays are tracked robustly, making no assumptions about small deflection angle or unperturbed geodesics, i.e.\ the Born approximation. 10,000 map realizations were generated for each model by rotating and shifting the spatial planes. Each map has 512$^2$ pixels and is 12.25~deg$^2$ in size. We refer the readers to the code paper for more simulation details and code testing.

The \texttt{LSST} Science Book forecasts a projected number density of 50~arcmin$^{-2}$ for the survey, with a redshift distribution that peaks at $z$=1 and 10\% of galaxies at $z > 2.5$ ~\citep{lsstbook}. We adopt a five redshift tomography setting, assuming a galaxy number density $n_{\rm gal}$=8.83, 13.25, 11.15, 7.36, 4.26~arcmin$^{-2}$ for source redshifts $z_s$=0.5, 1, 1.5, 2, 2.5, respectively. To demonstrate the power of tomography~\citep{Hu2002}, we also study the constraints from one single redshift, where all galaxies are at $z_s=1.0$ and $n_{\rm gal}=40$~arcmin$^{-2}$. 
For both configurations, we considered galaxy shape noise $\sigma_{\lambda} = 0.35$. 

To model the \texttt{LSST} noise, we add $\kappa_{\rm noise}$ to each pixel, drawn from a random Gaussian distribution centered at zero with variance
\begin{equation}\label{eq:noise}
\sigma^2_{\rm noise} = \frac{ \langle \sigma^2_{\lambda} \rangle }{n_{\rm gal} \Delta \Omega}.
\end{equation}
The noise level depends on the galaxy shape noise, the galaxy density, and the smoothing scale. For our default smoothing scale of two arcmin, $\sigma_{\rm noise}$=[0.0169, 0.0138, 0.0150, 0.0185, 0.0243] for the redshift bins $z_s$=[0.5, 1, 1.5, 2, 2.5], respectively. The noise level for the single redshift case is $\sigma_{\rm noise}$=0.0079.

Finally, we measure both the power spectrum and peak counts for each map. For the power spectrum, we use logarithmically spaced bins with neighboring bin edges increasing by a factor of 1.124, with $\ell_{\rm min}$=100 and $\ell_{\rm max}$=5000 (equivalent to $\sim$2 arcmin in real space). For peak counts, we use linearly spaced bins between S/N=$-0.6$ to 6 with $\Delta S/N$=0.16, where the noise term is defined in Eq.~\ref{eq:noise}. 

\section{\label{sec:analysis}Analysis}

\subsection{Interpolation with Gaussian Process}

\begin{figure}
      \includegraphics[width=\columnwidth]{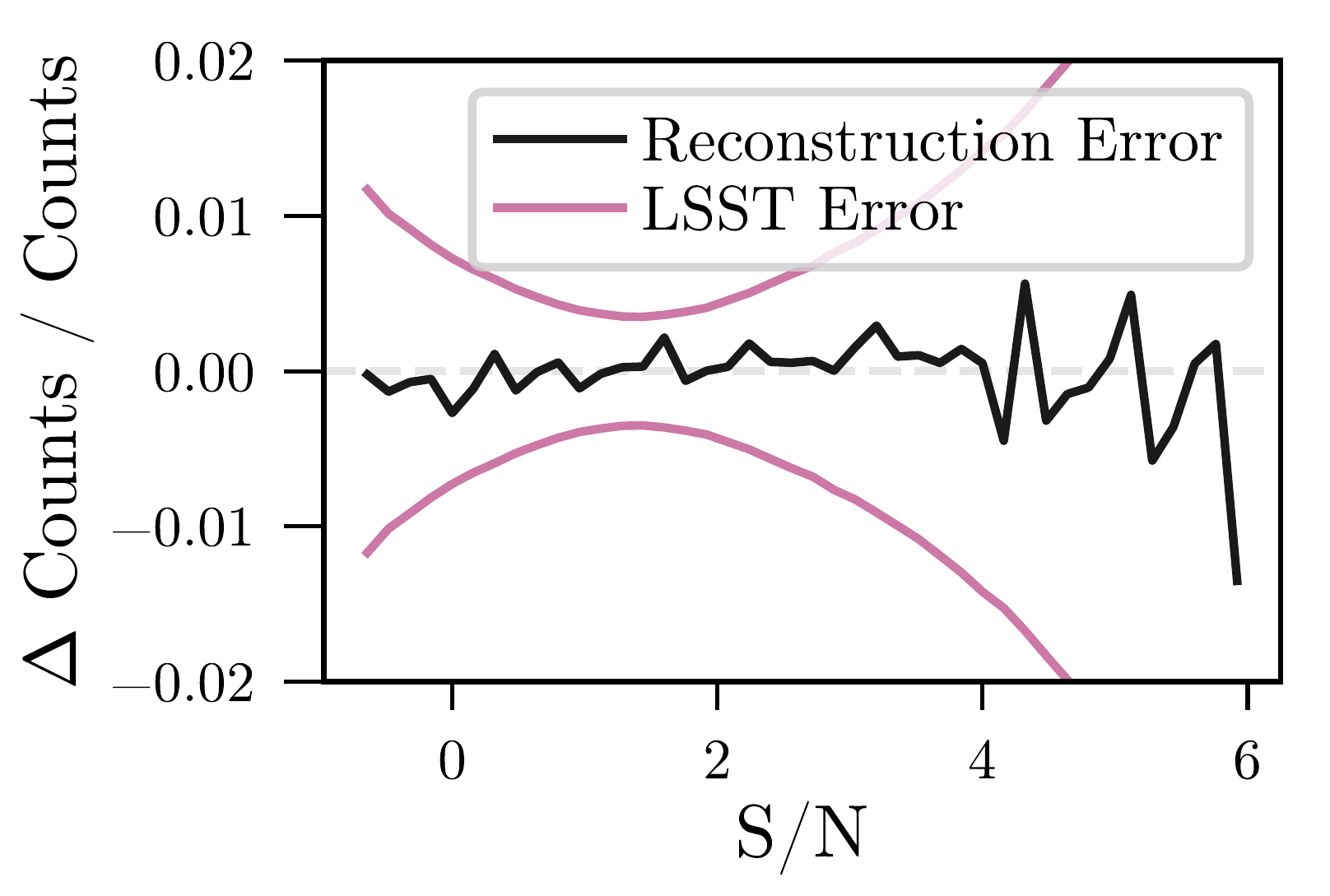}
      \includegraphics[width=\columnwidth]{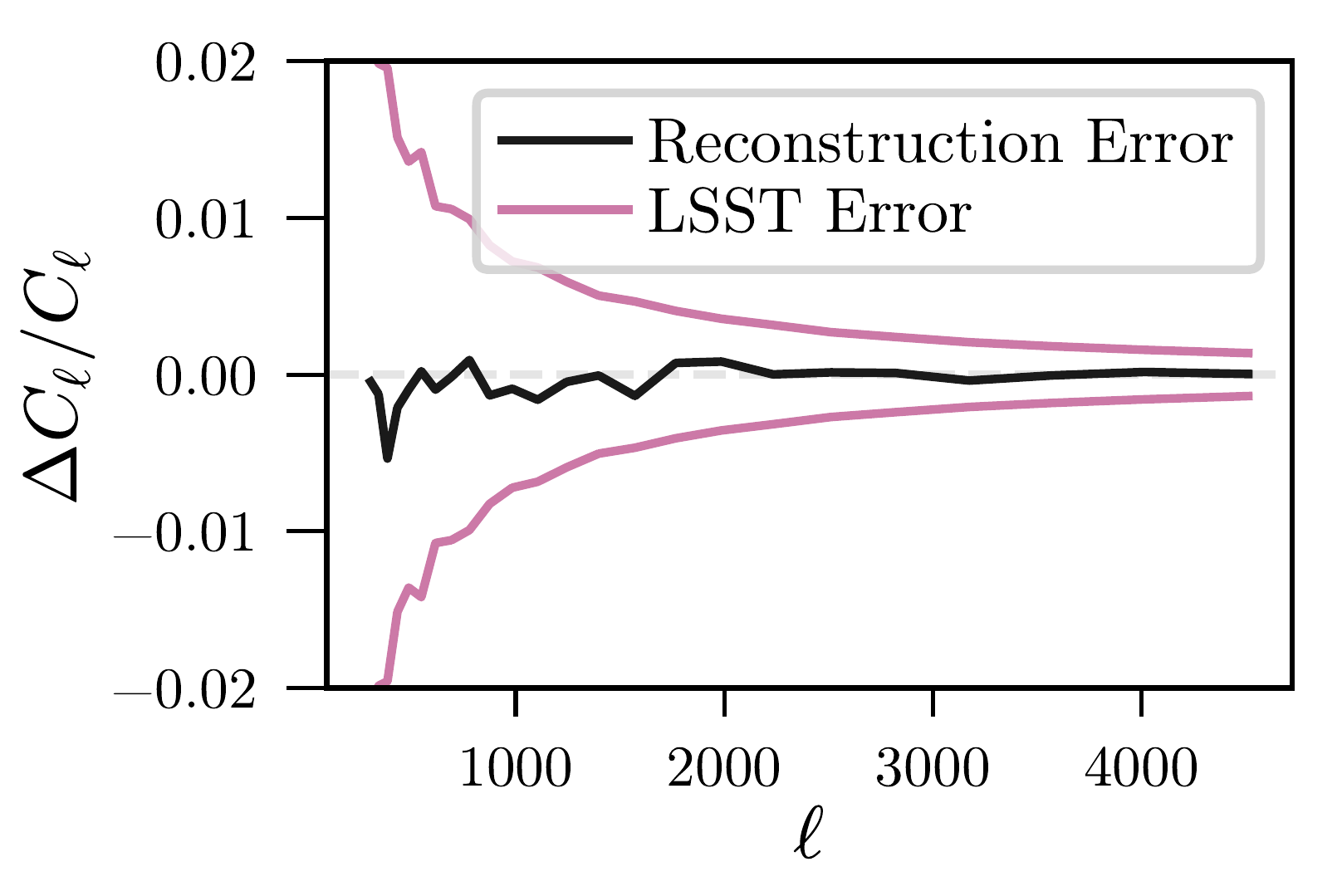}
      
  \caption{\label{fig:reconstruction_peaks_PS}The fractional interpolation error from the Gaussian Process (GP) for peak counts ({\bf top}, black) and power spectrum ({\bf bottom}, black). We test the robustness of the interpolation by comparing the true statistic at the fiducial model with that obtained from our GP interpolator (built with the other 100 models), We also show the expected variance of an \texttt{LSST}-like survey (purple).}
\end{figure}

To build models for the power spectrum and peak counts at an arbitrary cosmology, we interpolate between the available cosmological models~(Fig.~\ref{fig:sim_parameter_distribution}) 
using Gaussian Process (GP). 
Unlike a simple spline interpolator, which only uses the average information, GP also uses the variance to weight different bins and cosmologies when performing the interpolation. 
We find roughly an order of magnitude improvement in comparison to the RBF scheme used in previous work~\citep{liu2016}. We fit our GP with an anisotropic squared-exponential kernel with four hyperparameters---an interpolation length scale for each of the three parameters, as well as an overall amplitude. These are fit with the standard marginal likelihood approach \citep{gp2006} implemented in \texttt{scikit-learn} \citep{scikit-learn}.

We assess the reconstruction accuracy by comparing the GP prediction with the ``ground truth'' from the simulation. Here, the test GP is built with that particular model removed. We found consistently sub-percent errors, well below the systematical error expected from \texttt{LSST}~\footnote{We also test the improvement on interpolation by including the $\ell$ or $\kappa$ as an input parameter, i.e.\ fitting the GP on simulation data varying ($\ell, \sum m_{\nu}, \Omega_m, A_s$) or  ($\kappa, \sum m_{\nu}, \Omega_m, A_s$), as well as the redshift, though found negligible effects.}. We show an example for the fiducial model in Fig.~\ref{fig:reconstruction_peaks_PS}.

\subsection{Covariance}

\begin{figure*}
  \includegraphics[width=\columnwidth]{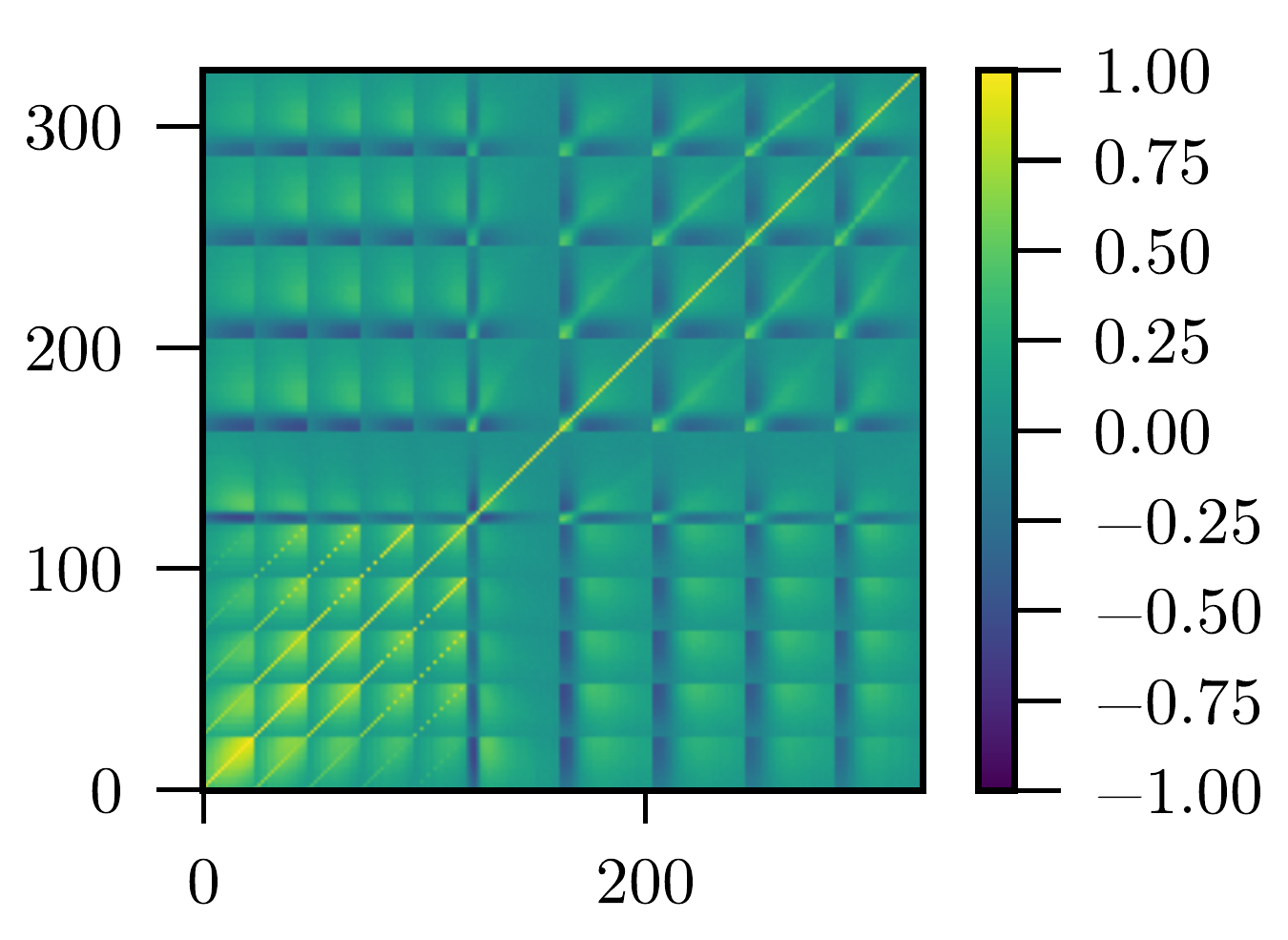}     \includegraphics[width=\columnwidth]{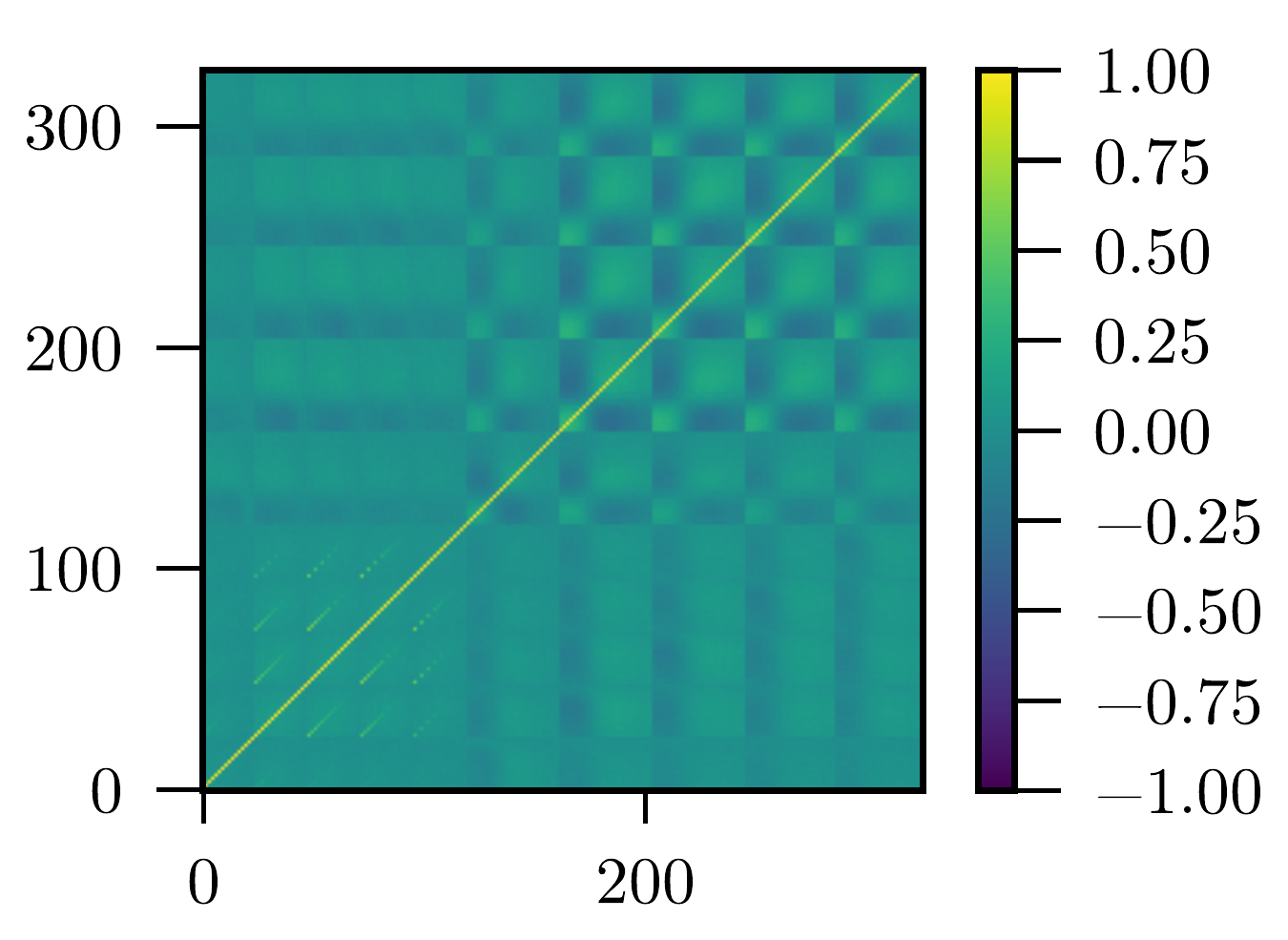}

  \caption{\label{fig:cor}Noiseless ({\bf left}) and noisy ({\bf right}) covariance matrices, normalized by the diagonal terms, for the power spectrum (bins 1--120) and peak counts (bins 121--326). We assume \texttt{LSST} noise~(Sec.~\ref{sec:simulations}). Five source redshifts ($z_s$=0.5, 1, 1.5, 2, 2.5) are shown as the five mini-blocks within each statistic.} 
\end{figure*}

We estimate the covariance matrices using 10,000 realizations from a separate, independent set of simulations at the fiducial massless model ($\sum m_{\nu} = 0$ eV, $\Omega_m = 0.3$, $A_s = 2.1 \times 10^{-9}$), to avoid correlations between the model and the covariance. We conduct separate and joint analyses for the power spectrum and peak counts in this paper. When the two statistics are combined, the full covariance is used. 

We show the full covariance for both the noiseless and noisy power spectrum (bins 1--120) and peak counts (bins 121--326) in Fig.~\ref{fig:cor}. Within each of the two main blocks, five sub-blocks are for the tomographic redshift bins ($z_s$=0.5, 1, 1.5, 2, 2.5, from left to right, respectively). For the power spectrum, larger nonlinear signals are seen at lower redshifts as larger values in the off-diagonal terms. The cross blocks between the power spectrum and peak counts show amplitude smaller than the self off-diagonal terms of the power spectrum, hinting that peak counts is a probe relatively independent from the power spectrum.

The inverse covariance is corrected for the limited number of realizations used~\citep{hartlap2007}, 
\begin{equation}
{C}^{-1} = \frac{N_{\rm s}-N_{\rm p}-2}{N_{\rm s}-1} {C}^{-1}_*,
\end{equation}
where $N_{\rm s}$ is the number of realizations, $N_{\rm p}$ is the number of bins, ${C}$ is the corrected covariance, and ${C}_*$ is the covariance computed from the simulations. Our simulated lensing maps are 12.25~deg$^2$, but \texttt{LSST} covers roughly 2$\times$10$^4$ deg$^2$. Therefore, we scale our covariance by $12.25 / (2 \times 10^4)$ to account for this difference in sky coverage.

\subsection{Likelihoods and parameter estimation}

We compute the likelihood assuming Gaussian errors, as the maps are Gaussian noise dominated. 
The log likelihood for a cosmology-independent covariance is
\begin{equation}
\log \mathcal{L}(p) = - \frac{1}{2} \left(x-\mu(p)\right)^T C^{-1} \left(x-\mu(p)\right),
\end{equation}
where $x$ is the vector of ``observed'' power spectrum or peak counts, for which we use the average from the fiducial cosmology ($\sum m_{\nu} = 0.1$ eV, $\Omega_m = 0.3$, $A_s = 2.1 \times 10^{-9}$), and $\mu$ is the GP prediction for the cosmological parameters~$p$. 
We estimate the posteriors of parameters using Markov Chain Monte Carlo (MCMC) with the \texttt{emcee} package, using 100 walkers initialized in a tight ball of radius $10^{-3}$ in each parameter about the fiducial model. 

\section{\label{sec:results}Results}

In this section, we present forecasts on the neutrino mass sum $\sum m_{\nu}$, the matter density $\Omega_m$, and the primordial power spectrum amplitude $A_s$ given a galaxy survey with \texttt{LSST} noise properties. $\Omega_m$ and $A_s$ are two parameters expected to be most degenerate with the neutrino mass for weak lensing. We compare the constraints derived from the lensing power spectrum and peak counts, as well as them jointly. We also present studies of varying the configurations: (1) single source redshift versus five tomographic redshifts; (2) the $\ell$ range included in the power spectrum analysis; (3) the smoothing scale for peak counts; (4) constraints from peaks of different heights (or significance); and (5) combining with primordial CMB priors. A summary of our findings can be found in Fig.~\ref{fig:megabar} and Table~\ref{tab:1Derror}.

\begin{figure*}      
\includegraphics[width=2\columnwidth]{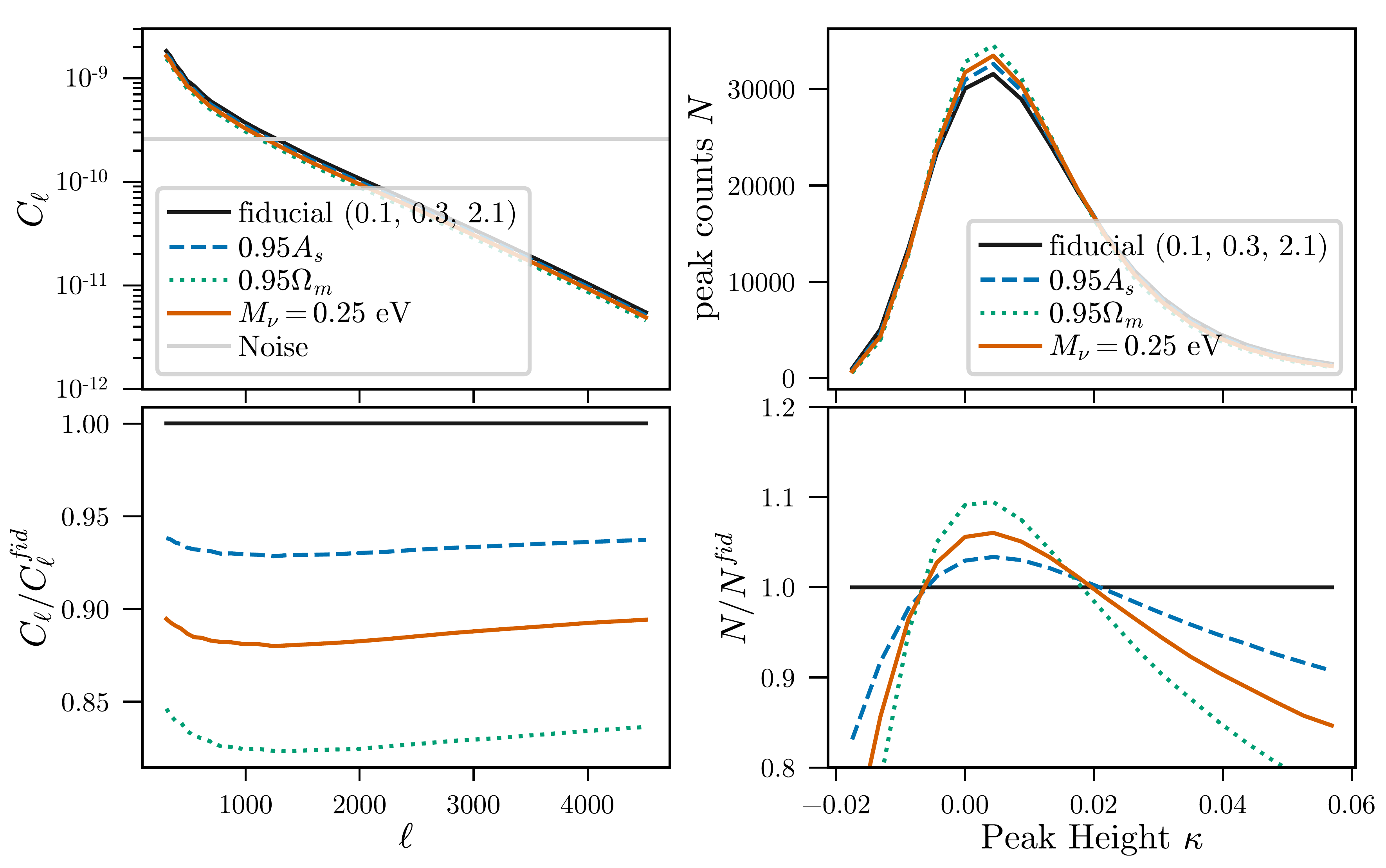}
      
  \caption{\label{fig:PS_Peak_effects}{\bf Left}: The impact of changes in the cosmological parameters on the noiseless lensing power spectrum at $z=1$, shown as fractional changes from the fiducial model (solid black, $\sum m_{\nu}$=0.1~eV, $\Omega_m$=0.3, $A_s$=2.1$\times 10^{-9}$). We generate the other three curves from our Gaussian Process interpolator, while keeping the other two parameters fixed.
  Decreasing $A_s$ or $\Omega_m$, and increasing $\sum m_{\nu}$ all decrease the overall lensing power spectrum, with very subtle changes in the shape of the curve, which explains the degeneracy among these parameters. {\bf Right}: same but for the peak counts. Similarly, the three parameters show degenerate behavior. However, a close examination of the bottom panel shows different crossing at $N=N_{fid}$ for the three parameters, hinting on potential to break the degeneracy. Also shown in the top left is the amplitude of white noise for $n_{gal} = 40$ per arcmin$^2$.}
\end{figure*}

\subsection{Power spectrum}

We show the impact of neutrino mass on the power spectrum in the left panel of Fig.~\ref{fig:PS_Peak_effects}. For comparison, we also vary the other two parameters, to demonstrate the well-know degeneracy among them---decreasing $A_s$ or $\Omega_m$, and increasing $\sum m_{\nu}$ all decrease the overall lensing power spectrum, with very subtle changes in the shape of the curve.

\begin{figure*}
      \includegraphics[width=\columnwidth]{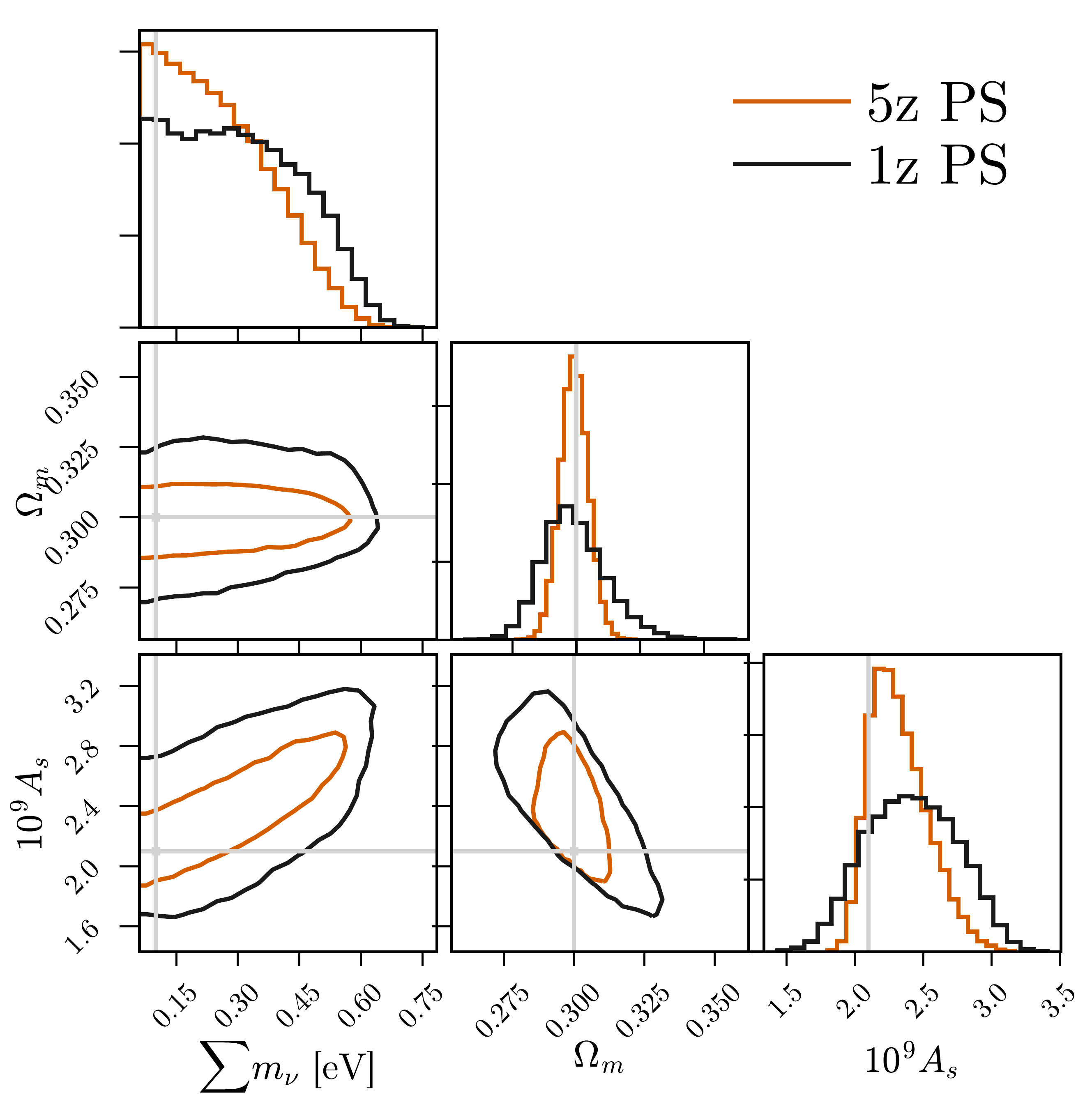}
      \includegraphics[width=\columnwidth]{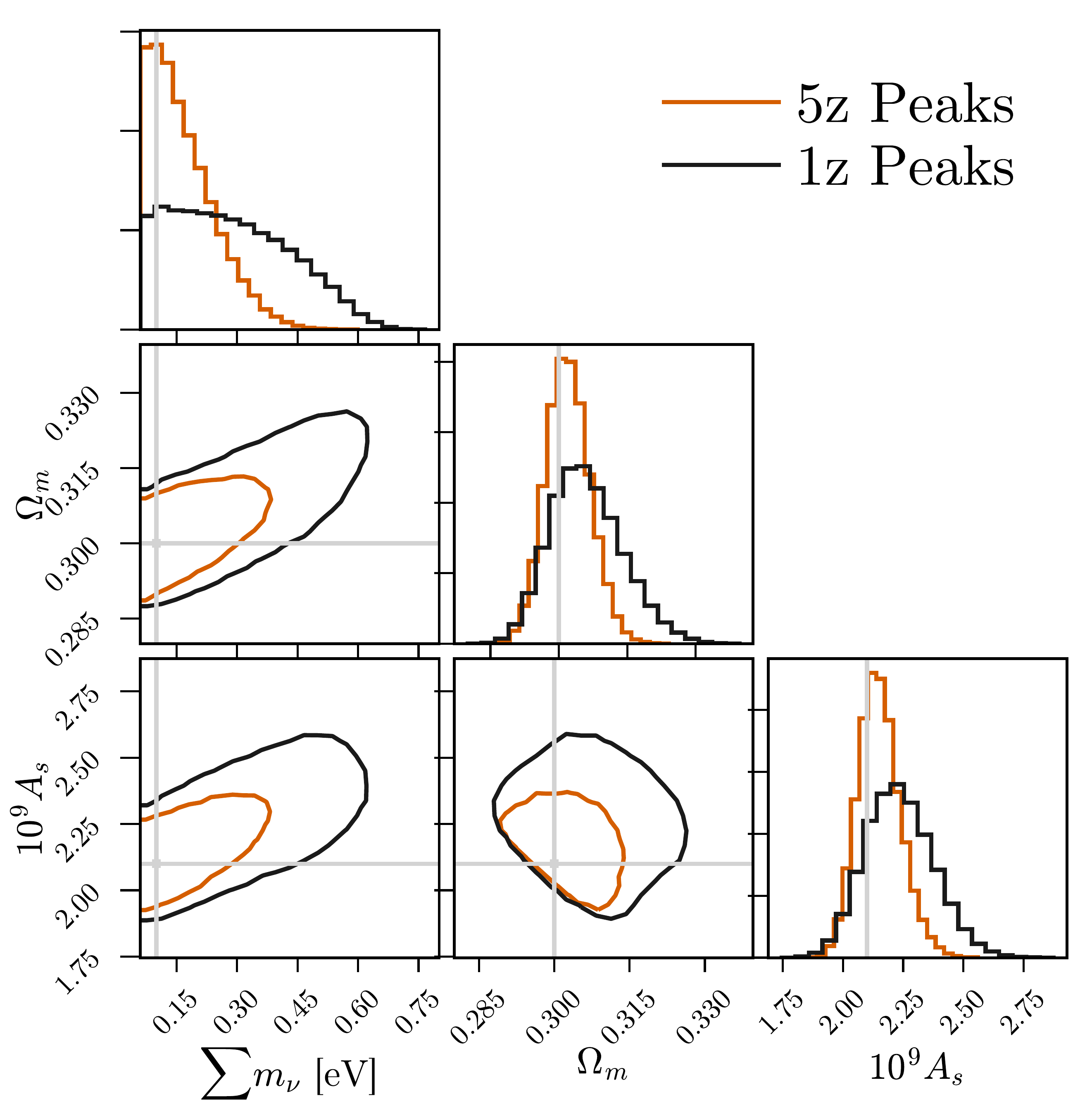}
  \caption{\label{fig:PS_tom}95\% confidence contours from the lensing power spectrum ({\bf left}) and peak counts ({\bf right}) assuming a single source plane at $z$=1 (black, ``$1z$'') and source planes across five redshifts $z$=\{0.5, 1, 1.5, 2, 2.5\} (orange, ``$5z$''), with total galaxy density held fixed. The fiducial values are shown in gray.}
\end{figure*}

We first examine the constraining power for a single source redshift (``$1z$'') versus five tomographic redshifts (``$5z$''), i.e.\ if we put all  galaxies in one single lensing map, or split them into different redshift bins for multiple maps. The motivation is to break the degeneracy among the three parameters. For example, while $A_s$ sets the initial power spectrum shape and hence only impacts the overall amplitude, $\Omega_m$ can change the shape of the power spectrum by shifting the matter-radiation equality, and $\Sigma m_\nu$ has a strong dependence on time as neutrinos gradually change from relativistic to non-relativistic. ~\citet{Hu2002} found that splitting galaxies to only a few tomographic bins can already gain up to an order of magnitude improvement. We show the results in  the left panel of Fig.~\ref{fig:PS_tom}. We use lensing multipoles $300 < \ell < 5000$ for both $1z$ and $5z$ configurations. Using $5z$ tomography, the constraints are improved by 12\%, 55\%, and 34\% for $\sum m_\nu, \Omega_m, and A_s$, respectively.

\begin{figure}
      \includegraphics[width=\columnwidth]{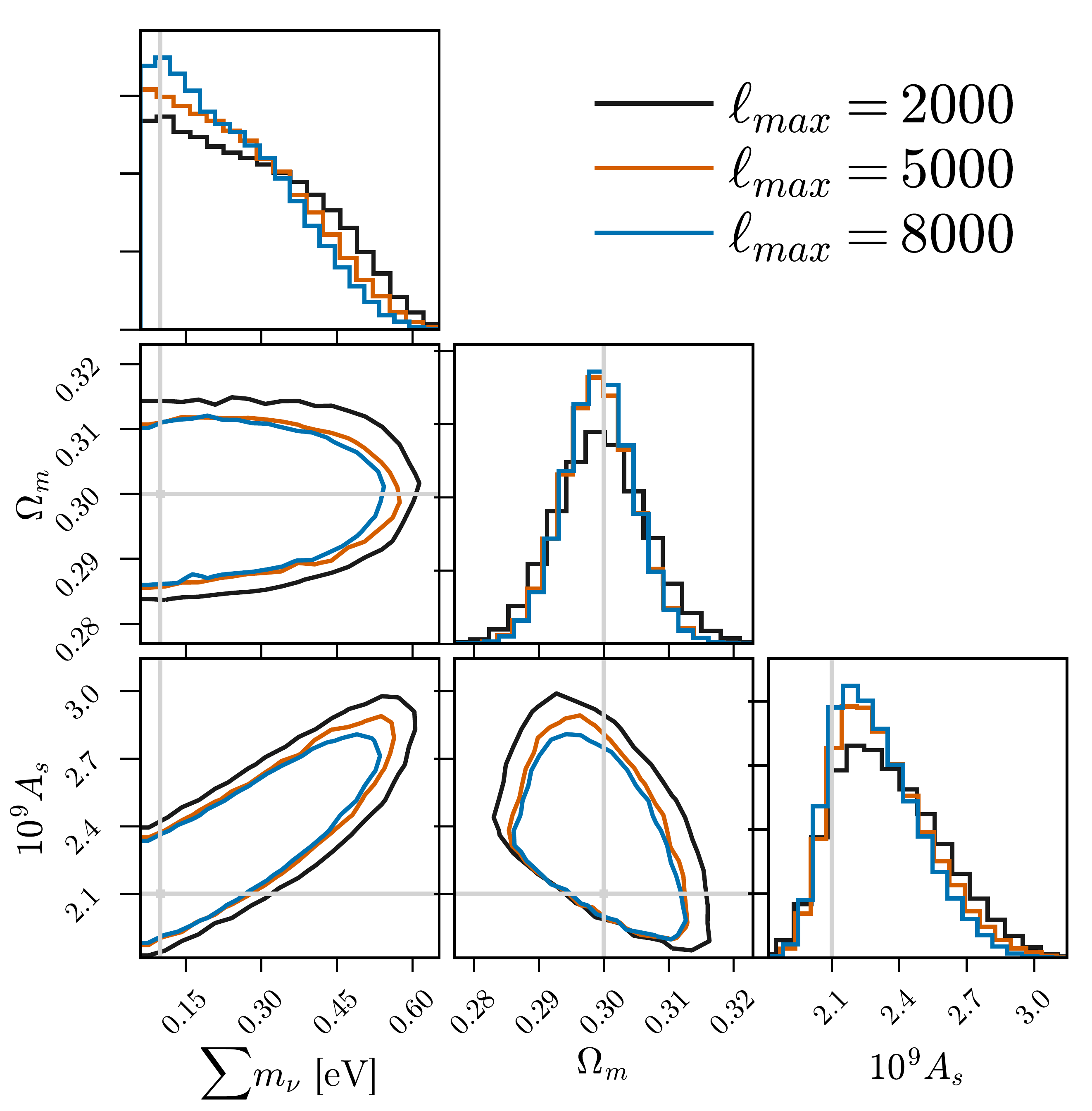}
  \caption{\label{fig:PS_ell}A comparison of cosmological parameter constraints from the lensing power spectrum for different maximum multipoles used in the analysis. The contours represent 95\% confidence contours taking information from multipoles ranging from $300 < \ell < \ell_{\rm max}$, with $\ell_{\rm max}$=2000 (black), $\ell_{\rm max}$=5000 (orange), and $\ell_{\rm max}$=8000 (blue). The fiducial values are shown in gray.} 
\end{figure}

In Fig.~\ref{fig:PS_ell} we compare the parameter constraints derived from different multipole cutoffs at $\ell_{\rm max}$=2000, 5000, 8000. In the noise-free case, we expect that including smaller scales (higher $\ell$) would add more information. Overall, the constraints are stable to changes in the maximum multipole, as expected. When we extend from $\ell_{\rm max}$=2000 to $\ell_{\rm max}$=5000, we see some mild improvements for all parameters.
However, pushing further to $\ell_{\rm max}$=8000 has negligible impact on the confidence levels, due to the dominance of galaxy noise at small scales (Fig.~\ref{fig:PS_Peak_effects}, left panel). 

\begin{figure}[h]
      \includegraphics[width=\columnwidth]{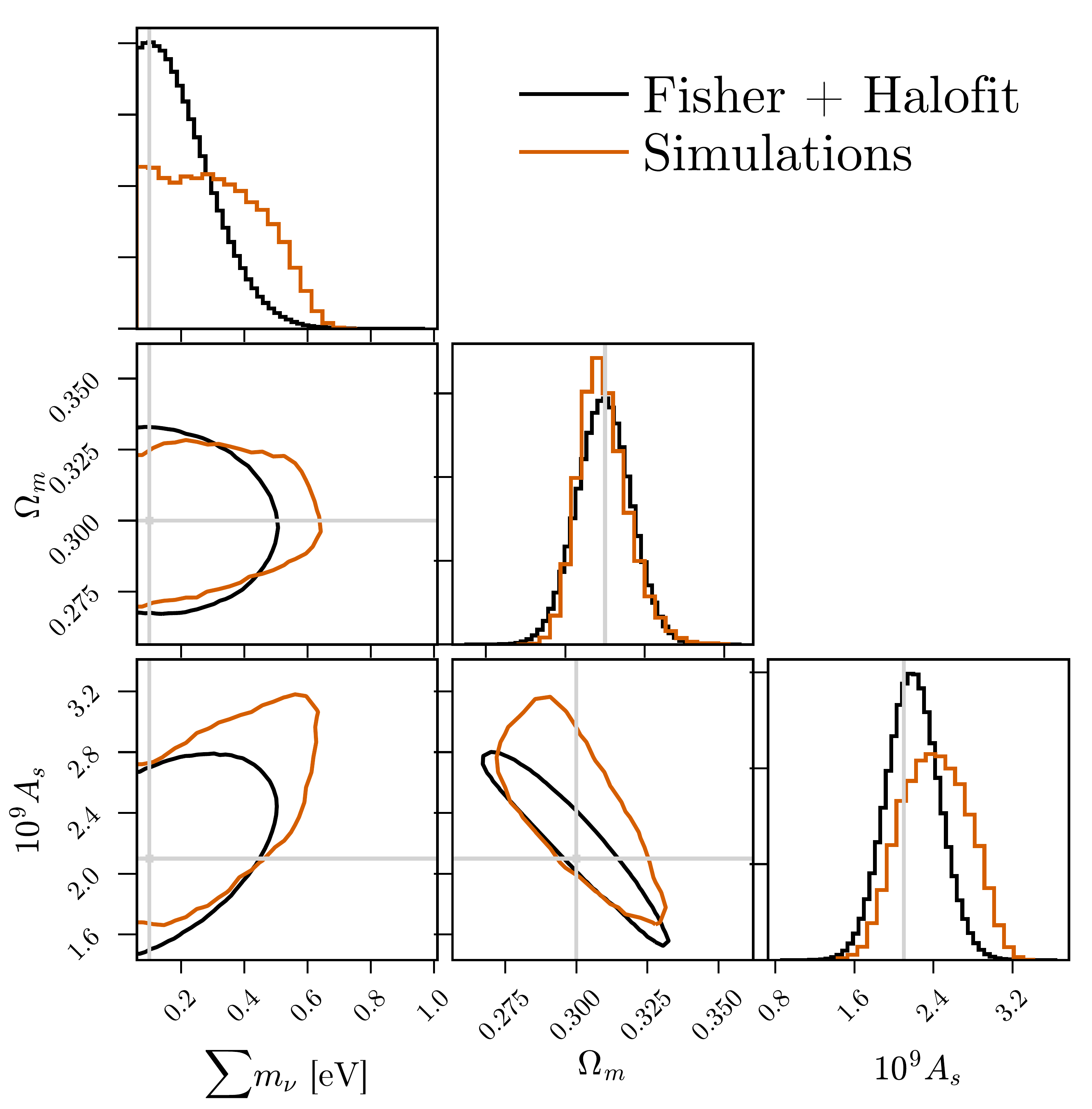}
  \caption{\label{fig:class}Comparison of the power spectrum 95\% confidence contours from our simulations (orange) to that from a simple Fisher forecast using Halofit (black).}
\end{figure}

Finally, in Fig.~\ref{fig:class} we check our simulation results against a simple Fisher forecast, using the lensing power spectrum predicted by the Halofit model~\citep{Smith2003,Takahashi2012}. We use a single redshift configuration (``1z''). The theory curve is computed using the CLASS software~\citep{Blas2011}. The degeneracy direction from the simulations agrees well with that from the Fisher forecast. The detailed contour shapes do not match exactly, though this is expected from the Fisher formalism, which only incorporates linear sensitivity to parameters, and Halofit, which is only accurate to 10\%. 

\subsection{Peak counts}

We show the impact of neutrino mass on lensing peaks in the right panel of Fig.~\ref{fig:PS_Peak_effects}. We also vary other two parameters to examine the degeneracy. Decreasing $A_s$ or $\Omega_m$, and increasing $\sum m_{\nu}$ suppress the number of low and high peaks, but boost the medium peaks. If we recall previous discussions that high peaks are found to be related to massive halos (Sec.~\ref{sec:background_peaks}), it is then not surprising to see that massive neutrinos reduce the number of massive halos in the universe.

We compare the parameter constraints obtained from a single source plane versus five redshifts for lensing peaks in the right panel of Fig.~\ref{fig:PS_tom}. Like the power spectrum, redshift tomography does provide additional information. Interestingly, whereas tomography for the power spectrum appears to uniformly shrink the contours in the $\sum m_{\nu}$--$\Omega_m$ plane, peak counts benefit from redshift information primarily in the direction of degeneracy. Using $5z$ tomography, the constraints are improved by 40\%, 39\%, and 36\% for $\sum m_\nu, \Omega_m, A_s$, respectively.

\begin{figure}
      \includegraphics[width=\columnwidth]{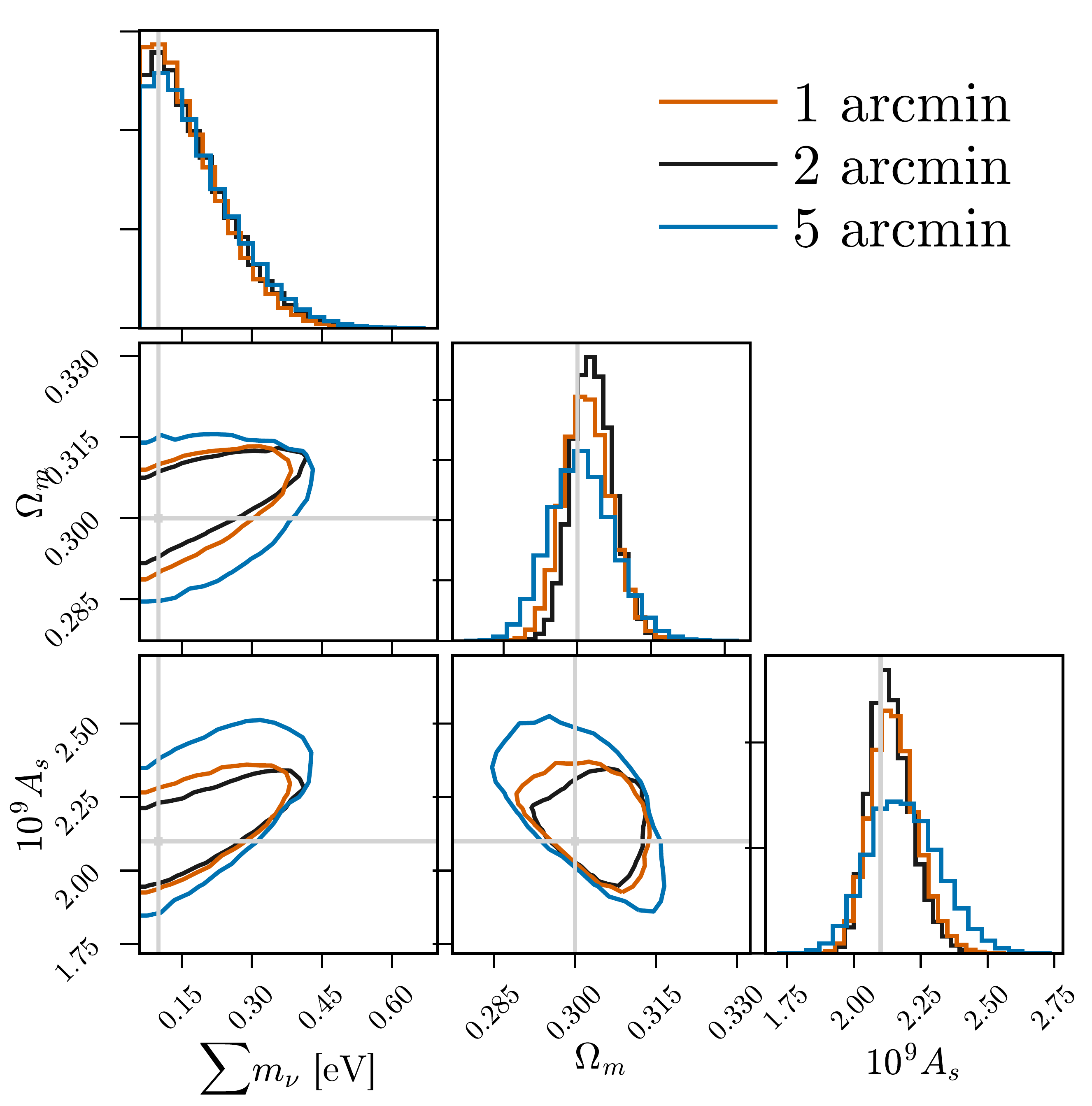}
  \caption{\label{fig:peaks_smoothing_scales}A comparison of 95\% confidence contours from lensing peak counts assuming different smoothing scales at 1 arcmin (orange), 2 arcmin (black) and 5 arcmin (blue). The fiducial values are shown in gray.}
\end{figure}

We also test the constraints from different smoothing scales in Fig.~\ref{fig:peaks_smoothing_scales}. While small smoothing scales can extract more nonlinear information, too small a scale would result in a noise dominated map. This is similar to the practice of applying an $\ell_{\rm max}$ to the power spectrum. We see some improvement from 5~arcmin to 2~arcmin smoothing, but almost no difference from 2~arcmin to 1~arcmin. Therefore, we choose to use 2~arcmin smoothing in our final analysis. We note that the constraints could be potentially improved further with more sophisticated filtering methods. For example, \citet{Liu2015} found that combining different smoothing scales can slightly shrink the contour size. \citet{Peel2018} discussed an alternative filtering scheme using wavelets, which can separate information on different scales more effectively than the Gaussian filter. 

\begin{figure}
  \includegraphics[width=\columnwidth]{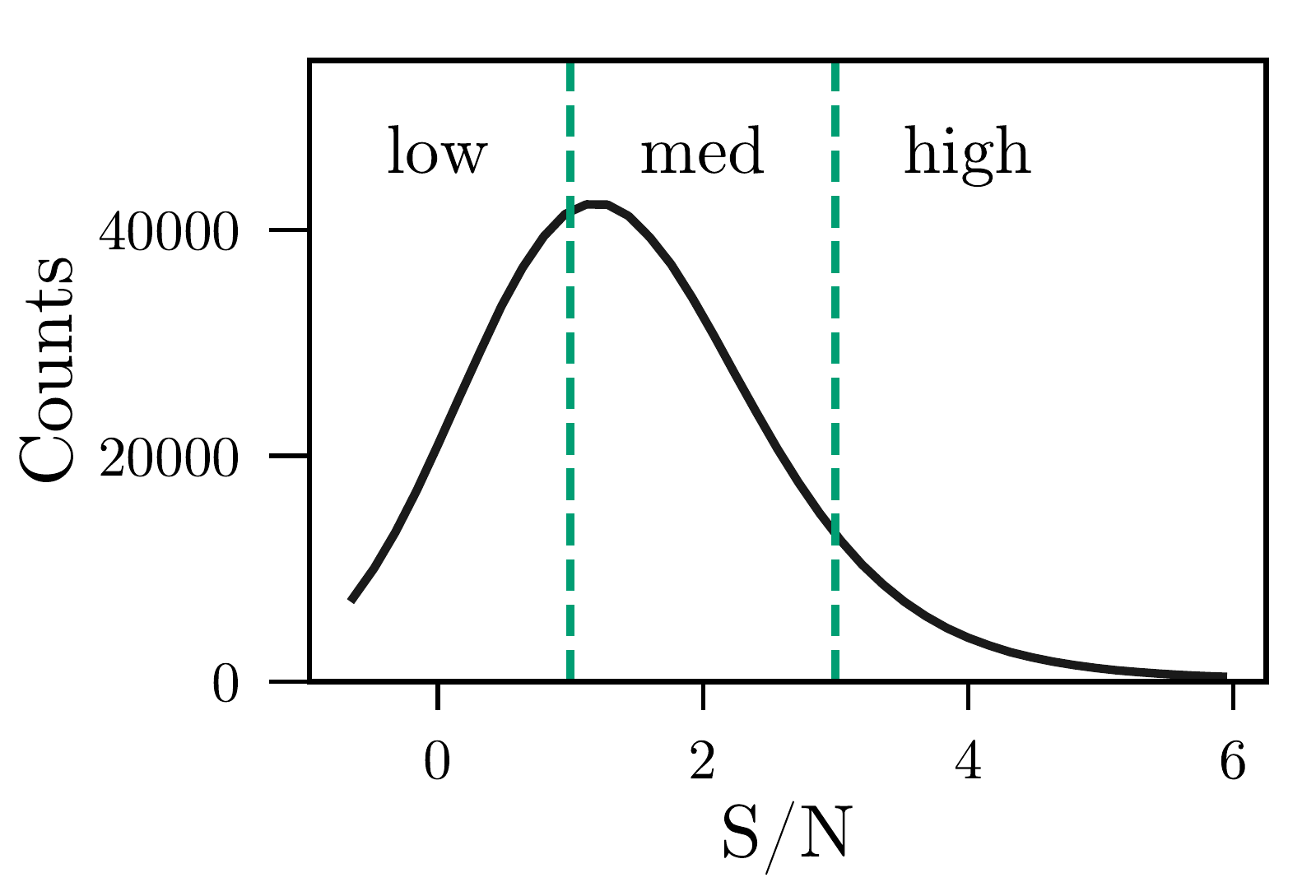}
  \caption{\label{fig:low_med_high}We split the peaks into three groups based on their height, separated by S/N=1 and S/N=3, as labeled  ``low'', ``med'', and ``high'' peaks. We show the single redshift peak count spectrum configuration ``1z'' with 2 arcmin smoothing scale for illustration.}
\end{figure}

\begin{figure}
      \includegraphics[width=\columnwidth]{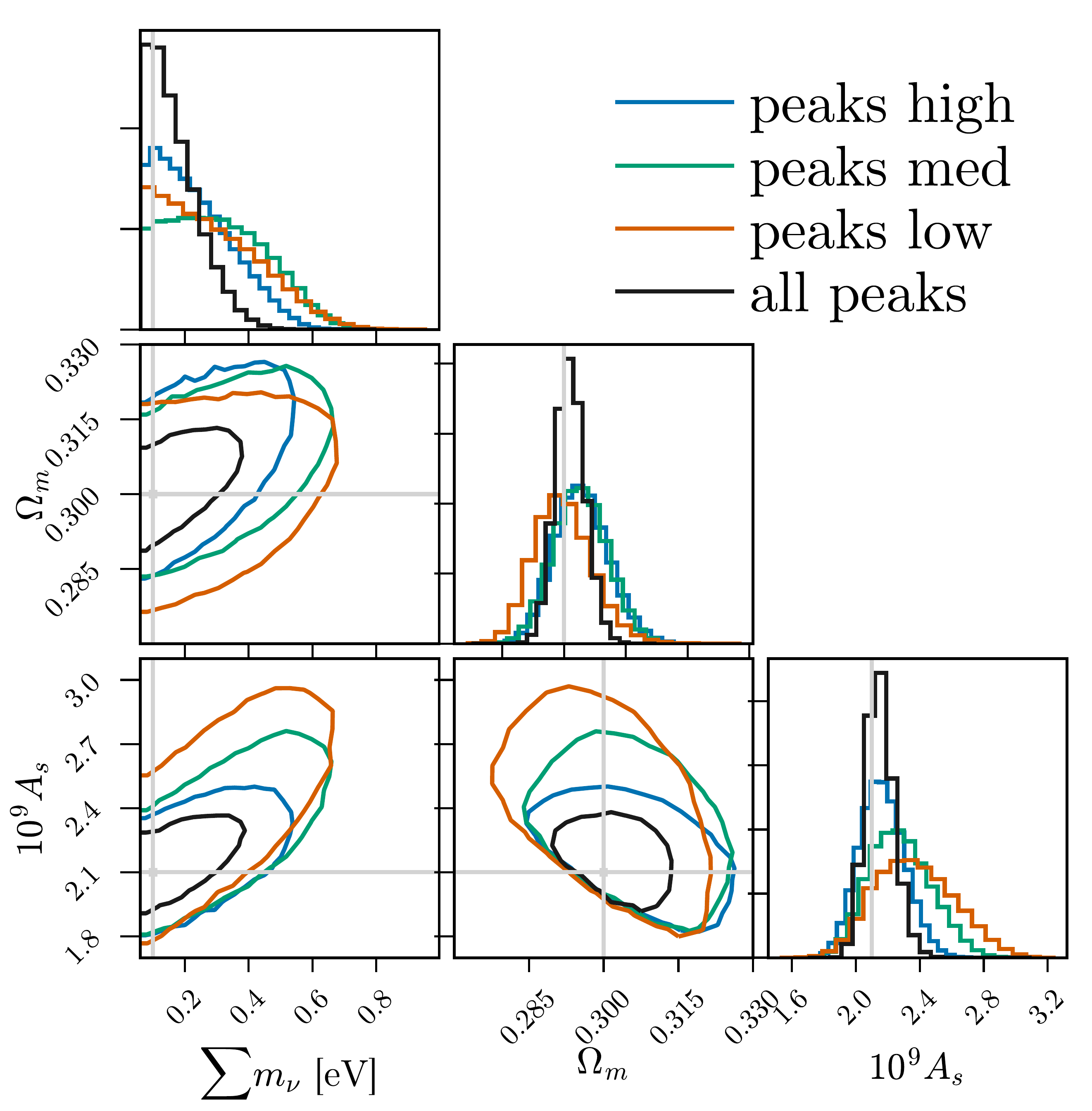}
  \caption{\label{fig:peak_bin}95\% confidence contours for the ``low'' (orange), ``med'' (blue), and ``high'' (black) peak height bins, as described in Fig.~\ref{fig:low_med_high}. The fiducial values are shown in gray.}
\end{figure}

Past work has suggested that much of the cosmological information in peak counts comes from the medium and low peaks. To study this, we separate the peaks according to their heights: low (S/N$< $1), medium (1$<$S/N$<$3), and high (S/N$>$3)  peaks. We illustrate our definitions in Fig.~\ref{fig:low_med_high}. We show the resulting parameter constraints in Fig.~\ref{fig:peak_bin}, where we see that medium and low peaks both contain similar amount of information as the high peaks. Peaks of different heights also show different degrees of~degeneracy.

\subsection{Joint constraints}

\begin{figure}
      \includegraphics[width=\columnwidth]{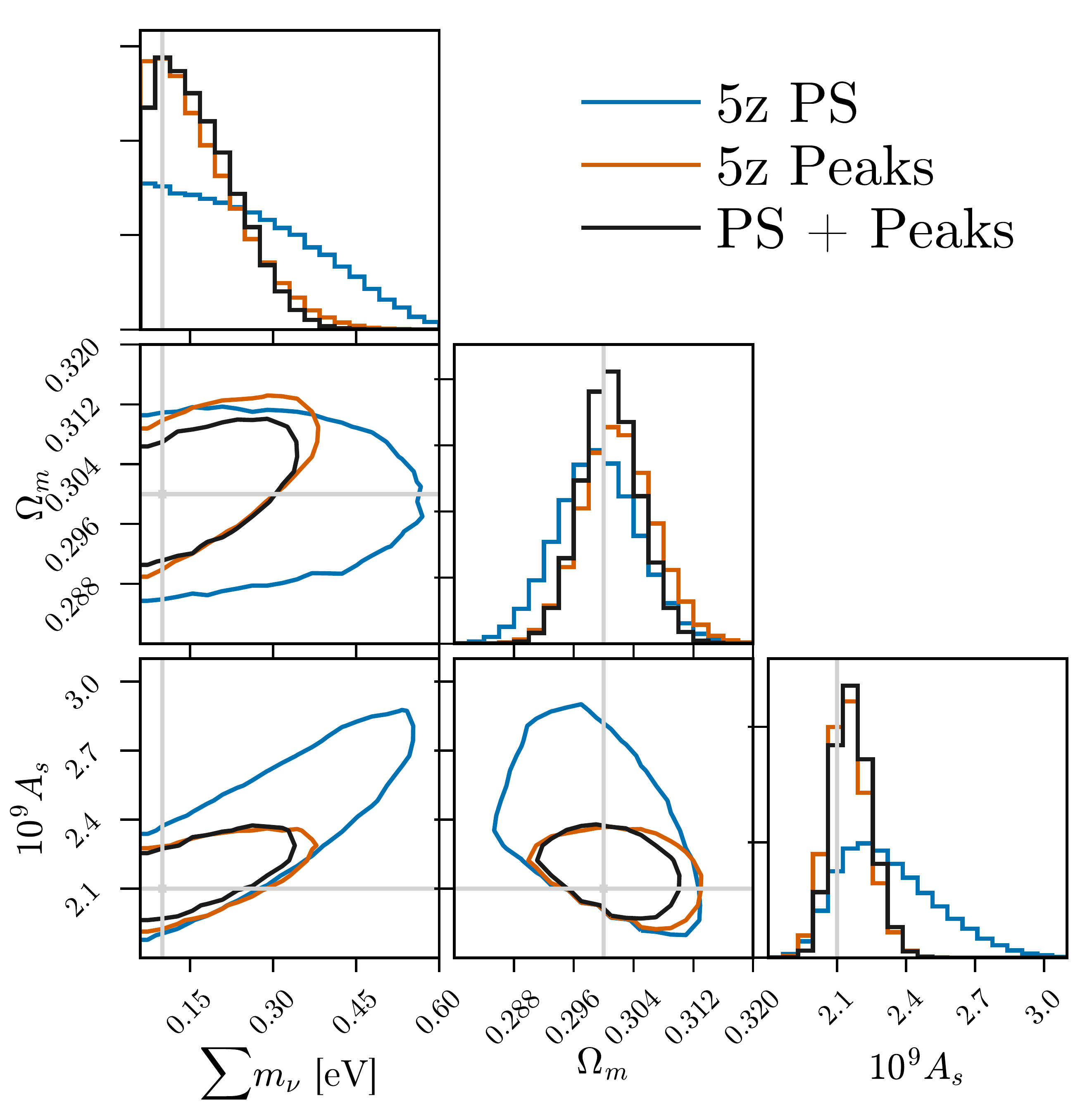}
  \caption{\label{fig:PS_vs_peaks}95\% confidence contours from the lensing power spectrum (``PS'',  $\ell_{\rm max}$=5000), peak counts (2~arcmin smoothing), and the two combined. We use five tomographic redshift bins. The full covariance is assumed. The fiducial values are shown in gray.}
\end{figure}

\begin{figure*}
      \includegraphics[width=1.8\columnwidth]{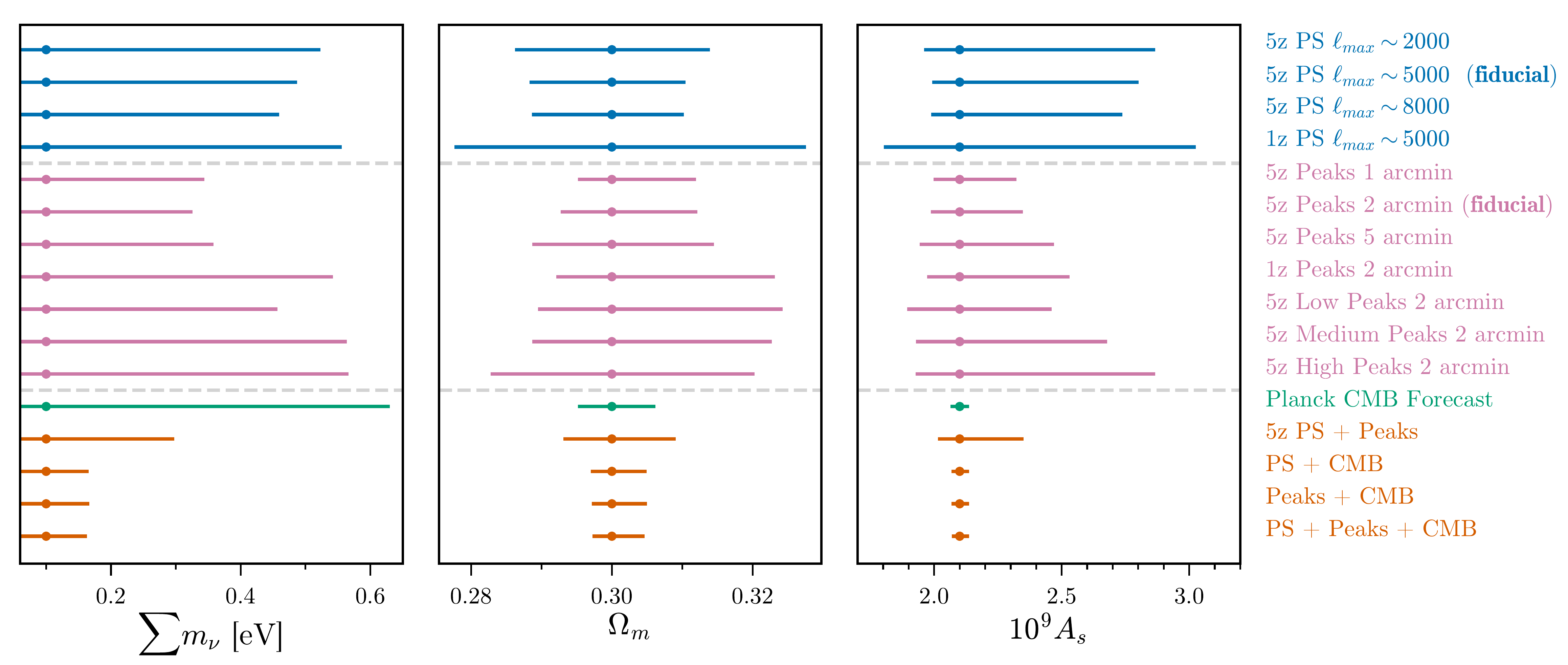}
  \caption{\label{fig:megabar}Marginalized constraints on each parameter for forecasts made in this paper, showing the 2.5 and 97.5 percentiles for $\Omega_m$ and $A_s$, as well as the 95\% upper bound on $\sum m_{\nu}$. The values are listed in Table~\ref{tab:1Derror}.}
\end{figure*}

We show the joint  95\% confidence level constraints in Fig.~\ref{fig:PS_vs_peaks}. 
We use a maximum multipole of $\ell_{\rm max}$=5000 for the power spectrum and a smoothing scale of 2 arcmin for peak counts. 
We also visualize the marginalized constraints in Fig.~\ref{fig:megabar}, with the values tabulated in Table~\ref{tab:1Derror}. Peak counts outperform the power spectrum in constraining all three parameters. The combined constraints are particularly impressive in $\sum m_{\nu}$, where the marginalized uncertainty is 39\% smaller, when compared to that from the power spectrum alone. 

\begin{figure}
      \includegraphics[width=\columnwidth]{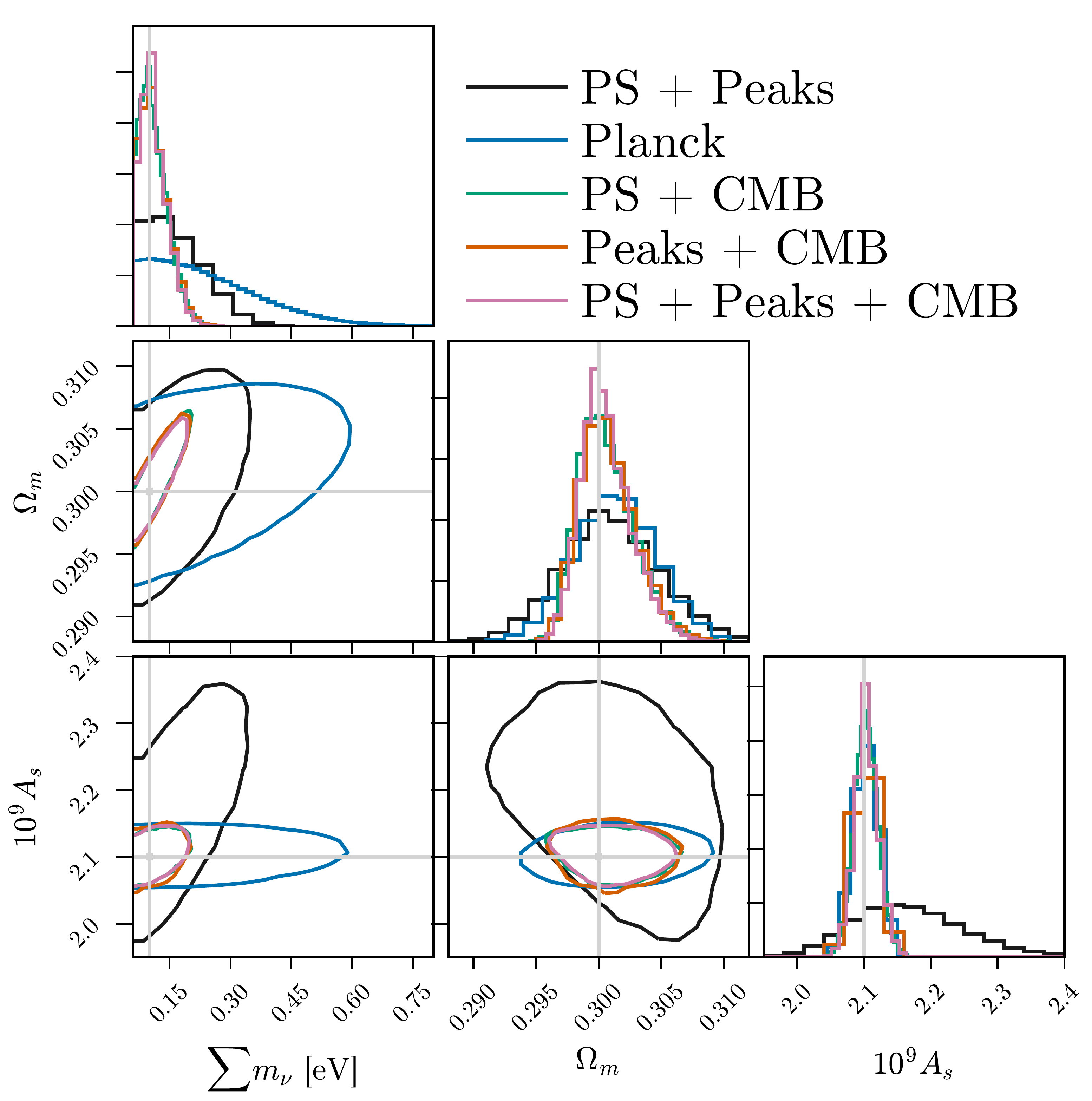}
  \caption{\label{fig:PS_vs_peaks_vs_cmb_combo}95\% confidence contours combined with the Planck prior (orange), the weak lensing power spectrum with five source planes (blue), and lensing peak counts with five source planes (black). The fiducial values are shown in gray.}
\end{figure}

The eventual strongest constraint on $\sum m_{\nu}$ would not be from weak lensing alone, but by combining with other  probes such as the CMB, baryon acoustic oscillation, Lyman-$\alpha$ forest, etc. We demonstrate this in Fig.~\ref{fig:PS_vs_peaks_vs_cmb_combo} with an example of combining our weak lensing likelihood with a Planck-like Fisher prior. 
While the CMB constraint on $\sum m_{\nu}$ is weaker than that from weak lensing, it is particularly sensitive to $A_s$, which breaks the strong degeneracy between $\sum m_{\nu}$ and $A_s$ in weak lensing, and hence resulting in a much tighter $\sum m_{\nu}$ than either probe alone. 

\begin{table*}
\begin{tabular}{|l|c|cc|cc|cc|}
\hline
observable & $\sum m_\nu+$ & $\Omega_m-$ & $\Omega_m+$ & $10^9A_s-$ & $10^9A_s+$ & $\Delta\Omega_m$ &   $\Delta 10^9A_s$ \\
\hline
$5z$ PS $\ell_{\rm max} \sim 2000$ & 0.520 & 0.286 & 0.314 & 1.968 & 2.859 & 0.027 & 0.892 \\
$5z$ PS $\ell_{\rm max} \sim 5000$ ({\bf fiducial}) & 0.484 & 0.289 & 0.310 & 1.999 & 2.795 & 0.022 & 0.796 \\
$5z$ PS $\ell_{\rm max} \sim 8000$ & 0.457 & 0.289 & 0.310 & 1.996 & 2.731 & 0.021 & 0.736 \\
$1z$ PS $\ell_{\rm max} \sim 5000$ & 0.553 & 0.278 & 0.327 & 1.810 & 3.019 & 0.049 & 1.209 \\
\hline
$5z$ Peaks 1 arcmin & 0.341 & 0.295 & 0.312 & 2.004 & 2.316 & 0.016 & 0.312 \\
$5z$ Peaks 2 arcmin ({\bf fiducial}) & 0.323 & 0.293 & 0.312 & 1.993 & 2.341 & 0.019 & 0.348 \\
$5z$ Peaks 5 arcmin & 0.355 & 0.289 & 0.314 & 1.949 & 2.463 & 0.025 & 0.513 \\
$1z$ Peaks 2 arcmin & 0.540 & 0.292 & 0.323 & 1.979 & 2.524 & 0.031 & 0.545 \\
$5z$ Low Peaks 2 arcmin & 0.454 & 0.290 & 0.324 & 1.901 & 2.454 & 0.034 & 0.553 \\
$5z$ Medium Peaks 2 arcmin & 0.561 & 0.289 & 0.322 & 1.936 & 2.672 & 0.034 & 0.736 \\
$5z$ High Peaks 2 arcmin & 0.564 & 0.283 & 0.320 & 1.934 & 2.860 & 0.037 & 0.926 \\
\hline
Planck CMB Forecast & 0.627 & 0.295 & 0.306 & 2.071 & 2.130 & 0.010 & 0.059 \\
5z PS + Peaks & 0.295 & 0.293 & 0.309 & 2.021 & 2.344 & 0.015 & 0.322 \\
PS + CMB & 0.163 & 0.297 & 0.305 & 2.075 & 2.130 & 0.007 & 0.055 \\
Peaks + CMB & 0.164 & 0.297 & 0.305 & 2.075 & 2.130 & 0.007 & 0.055 \\
PS + Peaks + CMB & 0.160 & 0.298 & 0.304 & 2.076 & 2.130 & 0.007 & 0.054 \\
\hline
\end{tabular}
\caption{\label{tab:1Derror}Marginalized constraints on each parameter for forecasts made in this paper, showing the 2.5 and 97.5 percentiles for $\Omega_m$ and $A_s$, as well as the 95\% upper bound on $\sum m_{\nu}$. The values are visualized in Fig.~\ref{fig:megabar}.}
\end{table*}

\section{\label{sec:conclusion}Conclusion}

In this paper, we study the constraints on the neutrino mass sum ($\sum m_{\nu}$) from the weak lensing power spectrum and peak counts for an \texttt{LSST}-like survey. We  study the effects of redshift tomography, $\ell_{\rm max}$ for the power spectrum, and smoothing scales for the peak counts. We also show the power of joint constraints of the power spectrum and peak counts, as well as that with primordial CMB temperature. 

We show the marginalized errors on $\sum m_{\nu}$, $\Omega_m$, and $A_s$ in Fig.~\ref{fig:megabar} for different combinations of the survey configurations discussed above. Our major findings are:

\begin{enumerate}
\item Redshift tomography improves parameter constraints for both the power spectrum and peak counts, compared with using only a single redshift bin. 
The marginalized constraints on the neutrino mass are improved by 12\% for the power spectrum, and 39\% for the peak counts (Fig.~\ref{fig:PS_tom}). 

\item Combining peak counts and the power spectrum can improve the constraint on $\sum m_{\nu}$ by $\sim$40\% over the power spectrum alone, as the two probes have different degrees of degeneracy between $\sum m_{\nu}$ and the other two parameters, $\Omega_m$ and $A_s$ (Fig.~\ref{fig:PS_vs_peaks}). It is worth noting that peak counts alone can already provide constraints on $\sum m_{\nu}$, $\Omega_m$, and $A_s$ that are competitive to the power spectrum. For example, the constraint on $\sum m_{\nu}$ from peaks alone is $\sim$33\% smaller than that from the power spectrum.

\item While including smaller scales in both the power spectrum and peak counts can improve the constraints, such effects are diminishing at around $\ell_{\rm max}=5000$ for the power spectrum and 2 arcmin smoothing for peak counts, due to the galaxy noise (Figs.~\ref{fig:PS_ell}~and~\ref{fig:peaks_smoothing_scales}). 

\item Low and medium peaks, typically formed due to multiple much smaller halos (than the single halos that cause the high peaks), contain a similar level of information as the high peaks, but with different degrees of degeneracy than the high peaks (Fig.~\ref{fig:low_med_high}). 

\item While weak lensing and CMB (with Planck noise) on their own provide comparable constraints on $\sum m_{\nu}$, when they are combined the constraint is improved by a factor of $\sim$4 compared to CMB alone (Fig.~\ref{fig:PS_vs_peaks_vs_cmb_combo}).

\end{enumerate}

In summary, we demonstrated that lensing peaks are a powerful tool on its own, probing nonlinear information that would be otherwise missed in the power spectrum analysis. The combination of the weak lensing power spectrum and peak counts tightens the constraint on neutrino mass for future galaxy surveys, than using the former alone, and is complementary to other cosmological probes. To eventually realize the power of peak counts in constraining neutrino mass, we must model carefully relevant systematics, including the intrinsic alignment, photometric redshift errors, galaxy shape bias, and baryonic feedback. We will investigate these in our future work.

\begin{acknowledgments}
We thank Jo Dunkley, Zoltan Haiman, Mathew Madhavacheril, David Spergel, and Ben Wandelt for helpful discussions. 
This work is supported by an NSF Astronomy and Astrophysics Postdoctoral Fellowship (to JL) under award AST-1602663. We acknowledge support from the WFIRST project. JZ was supported by NASA ATP grant 80NSSC18K1093. 
We thank New Mexico State University (USA) and Instituto de Astrofisica de Andalucia CSIC (Spain) for hosting the Skies \& universes site for cosmological simulation products.
This work used the Extreme Science and Engineering Discovery Environment (XSEDE), which is supported by NSF grant ACI-1053575. The analysis is in part performed at the TIGRESS high performance computer center at Princeton University. 
\end{acknowledgments}

\clearpage 
\bibliographystyle{aasjournal}
\bibliography{neutrino}

\end{document}